\documentclass[preprint]{ptephy_v1}

\preprintnumber{XXXX-XXXX} 
\usepackage{arydshln}
\usepackage{subcaption}
\usepackage{caption}
\usepackage{multirow}
\usepackage{arydshln}
\usepackage{amssymb}
\usepackage{amsmath}
\usepackage{url} 
\usepackage{xfrac}
\usepackage{lscape}
\usepackage{booktabs}

\begin{document}
\title{Development of a low-background HPGe detector at Kamioka Observatory}


\author[1, *]{K.~Ichimura}
\author[1]{H.~Ikeda}
\author[1,3]{Y.~Kishimoto}
\author[1]{M.~Kurasawa}
\author[1]{A.A.~Suzuki}
\author[1,6]{Y.~Gando}
\author[2,3]{M.~Ikeda}
\author[2]{K.~Hosokawa}
\author[2,3]{H.~Sekiya}
\author[4]{H.~Ito}
\author[5]{A.~Minamino}
\author[5]{S.~Suzuki}

\affil[1]{Research Center for Neutrino Science, Tohoku University, 6-3 Aoba, Aramaki, Aoba-ku, Sendai, Miyagi 980-8578, Japan}
\affil[2]{Kamioka Observatory, Institute for Cosmic Ray Research, The University of Tokyo, 456 Higashi-Mozumi, Kamioka, Hida, Gifu 506-1205, Japan}
\affil[3]{Kavli Institute for the Physics and Mathematics of the Universe (WPI), The University of Tokyo Institutes for Advanced Study, The University of Tokyo, 5-1-5 Kashiwanoha, Kashiwa, Chiba 277-8582, Japan}
\affil[4]{Department of Physics and Astronomy, Faculty of Science and Technology, Tokyo University of Science, 2641 Yamazaki, Noda, Chiba 278-8510, Japan}
\affil[5]{Department of Physics, Yokohama National University, 79-5 Tokiwadai, Hodogaya-ku, Yokohama, Kanagawa, 240-8501, Japan}
\affil[6]{Department of Human Science, Obihiro University of Agriculture and Veterinary Medicine, Inada-cho Nishi 2-11, Obihiro, Hokkaido 080-8555, Japan}
\affil[*]{\rm E-mail: ichimura@awa.tohoku.ac.jp}

\begin{abstract}
A new ultra-low background high-purity germanium (HPGe) detector has been installed at the Kamioka underground experimental site. 
The background count rate in the energy range from 40 keV to 2700 keV is about 25\% lower than that of the first HPGe detector installed in 2016, which has the same detector specification and similar shielding geometry.
This paper describes the shielding configuration, including the cleaning of the material surface, the comparison of calibration data and simulation, the time variation of the background spectra, the sample measurement procedure, and some results of the radioactivity in the selected samples. 
\end{abstract}

\subjectindex{H20}

\maketitle

\section{Introduction}
Material screening with high sensitivity is one of the important tasks for the current and next-generation rare event search  experiments.
For example, the SK-Gd project~\cite{Beacom:2003nk} requires that the concentration of radioactive impurities in Gd$_2$(SO$_4$)$_3{\cdot}$8H$_2$O should be less than  0.5~mBq~kg$^{-1}$ of the latter segment of U-chain (${}^{226}$Ra) and 0.05~mBq~kg$^{-1}$ of the latter segment of the Th-chain (${}^{228}$Ra) for the precise measurements of the solar neutrino spectrum~\cite{10.1093/ptep/ptac170}.
The KamLAND-Zen experiment~\cite{PhysRevLett.130.051801}, aiming to explore the neutrinoless double beta decay of $^{136}$Xe, plans to use a scintillation film and a wavelength shifter to enhance the sensitivity~\cite{10.1093/ptep/ptz064}. 
The radioactivities of the scintillation film and the wavelength shifter are required to be equal to or less than $\mathcal{O}$(1~ppt) for ${}^{238}$U and ${}^{232}$Th.
In order to measure radioactivity at the aforementioned required level, we developed the first ultra-low background (BG) high-purity germanium (HPGe) detector in 2016, hereafter referred to as Ge01.
This detector has been used to screen materials with high sensitivity~\cite{10.1093/ptep/ptac170,10.1093/ptep/pty096, 10.1093/ptep/ptaa105,Abe_2020,10.1093/ptep/ptaa119}. Then in 2020, we developed a new ultra-low BG HPGe detector with more careful treatment of the shielding material, to improve the sensitivity for increasing the capacity of material screenings for the second Gd-loading to Super-Kamiokande for 0.03\% Gd concentration~[K. Abe~{\it et al.}, manuscript in preparation] and for future projects conducted at Kamioka.\par
This paper describes the performance of the newly installed HPGe detector, including its analysis procedure for measuring radioactivity.
Sect.~\ref{sec:GeDetector} provides details of the detector configuration, data acquisition system, and calibration results.
Sect.~\ref{sec:Analysis} describes the data analysis, the time variation of the BG spectra, the comparison of the BG count rate with Ge01, and the sample measurement procedure.
A summary is presented in Sect.~\ref{sec:Conclusion}.

\section{Germanium Detector}
\label{sec:GeDetector}
\subsection{Detector configuration}
\label{sec:DetectorConfiguration}

\begin{figure}[htbp]
\centering
\includegraphics[width=0.8\textwidth]
{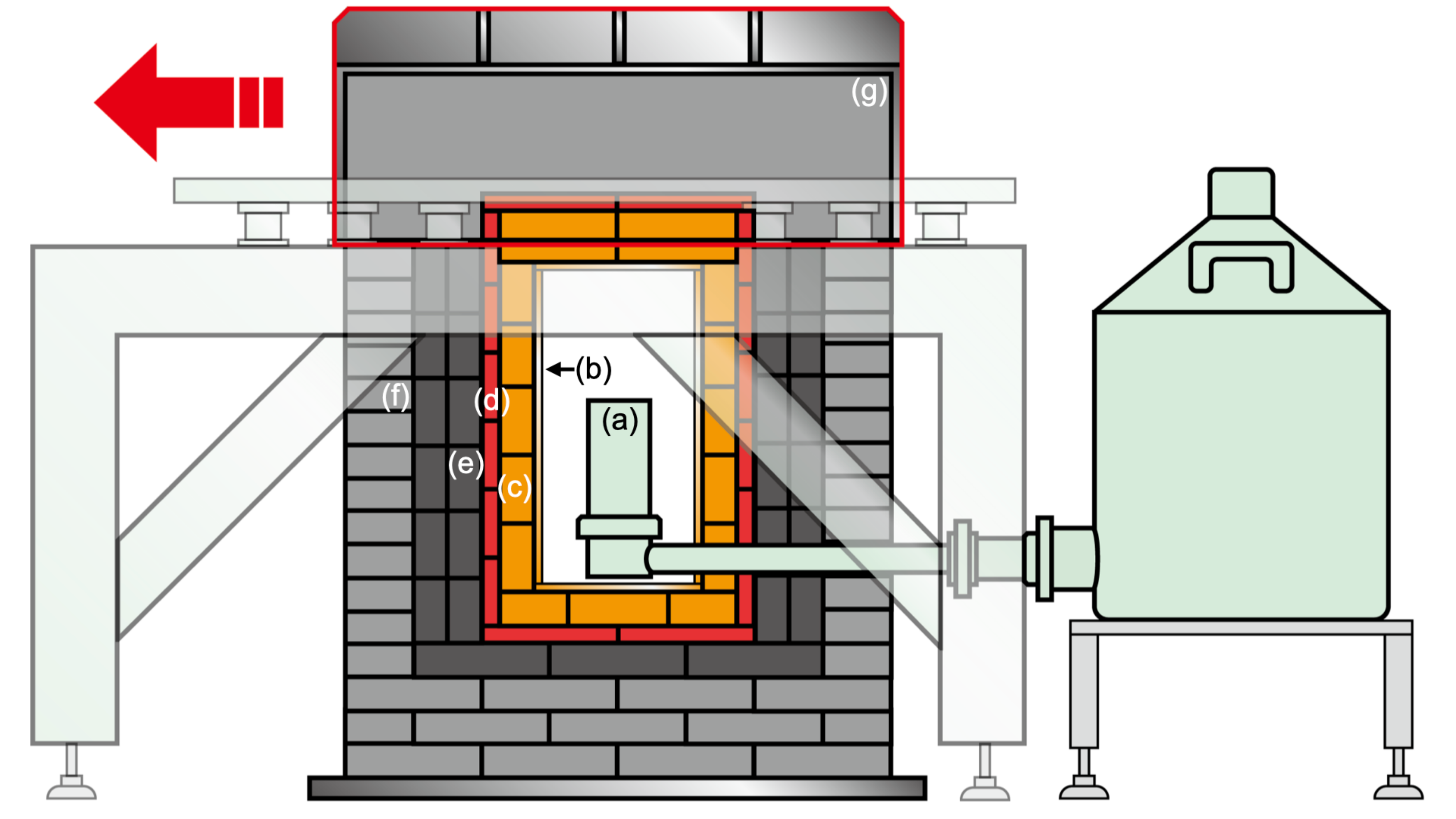}
\caption{Schematic view of the new HPGe detector. The HPGe detector with its cold finger and dewar (a), 6N grade copper (b), OFHC copper (c), lead layers with ${}^{210}$Pb activities of 5$\pm$3~Bq~kg$^{-1}$ (d), 35$\pm$4~Bq~kg$^{-1}$ (e), $\mathcal{O}(100)$~Bq~kg$^{-1}$ (f), and $\mathcal{O}(500)$~Bq~kg$^{-1}$ (g, only for the top part) can be seen. The top part circled with the red line can be floated and moved.}
\label{fig:LabCGePicture}
\end{figure}

The new HPGe detector, hereafter referred to as Ge02, has been manufactured by Mirion Technologies France~\cite{MirionTechnologies} with an ultra-low BG specification. 
Ge02 is a p-type coaxial detector with a crystal mass of 1.7 kg and a relative efficiency of 80\%\footnote{Efficiency is defined relative to a 3-inch diameter, 3-inch long NaI(Tl) crystal, for 1333 keV photopeak from a ${}^{60}$Co source positioned 25 cm away from the detector.}, the same specification as Ge01.
The detectors are installed at the Kamioka Observatory in the Kamioka Mine under Mt. Ikenoyama.
The average rock overburden is 2700 m water equivalent, which reduces the cosmic ray muon rate by 5 orders of magnitude.
Fig.~\ref{fig:LabCGePicture} illustrates the schematic view of Ge02.
The sample chamber has dimensions of 23$\times$23$\times$48~cm$^{3}$, with the HPGe detector positioned centrally within it.
The Ge crystal is housed in a cylindrical endcap made of ultra-low BG aluminum.
The spacious sample chamber allows the measurement of large sample quantities, such as 13$\times$3-inch photomultiplier tubes~\cite{Abe_2020} and 10~kg of Gd$_{2}$(SO$_{4}$)$_{3} \cdot$8H$_{2}$O~\cite{10.1093/ptep/ptac170}.
To mitigate the BG caused by radon (Rn) gas emanating from the shielding materials mentioned below, the sample chamber is continuously purged with Rn-free air at a rate of 5 standard liters per minute in which the concentration of ${}^{222}$Rn is $<1$~mBq/m$^{3}$~\cite{NAKANO2017108}.
The chamber is enveloped by 1~cm thick, 6N grade copper ($>$ 99.9999\% purity) supplied by Mitsubishi Materials Corporation~\cite{Mitsubishi}.
Surrounding the 6N grade copper is a 5~cm layer of oxygen-free-high-conductivity (OFHC) copper that has been stored at the Kamioka mine for over a decade. 
These layers of copper act as a shield against bremsstrahlung radiation generated by the beta-rays of ${}^{210}$Bi decay in the lead layers mentioned below.
The OFHC copper is surrounded by a 2.5~cm lead layer with a ${}^{210}$Pb activity of 5$\pm$3~Bq~kg$^{-1}$.
Then there is a 5~cm lead layer for the bottom (10~cm for the side) with a ${}^{210}$Pb activity of 35$\pm$4~Bq~kg$^{-1}$, purchased from SEIKO EG$\&$G Co Ltd.~\cite{SeikoEGandG}.
Outside of these low ${}^{210}$Pb lead layers are covered by lead with  a ${}^{210}$Pb activity of  $\mathcal{O}(100)$~Bq~kg$^{-1}$.

The upper part of the 5 cm~OFHC copper layer and the lead layers can be floated and moved laterally using a frame guide.
The innermost 2.5~cm of the lead layer in the top section has a ${}^{210}$Pb activity of 5$\pm$3~Bq~kg$^{-1}$ and the remaining 17.5~cm lead layer has that of $\mathcal{O}(500)$~Bq~kg$^{-1}$. 
In total, the lead layers have a combined thickness of 22.5~cm (20~cm for the top part) to shield against environmental gamma-rays.
Ge01 and Ge02 have the same thickness of the Cu layer and the combined lead layers. 
The difference in the shield configuration between Ge01 and Ge02 is that Ge01 has a ${}^{210}$Pb activity of O(100) Bq~kg$^{-1}$ lead for the layer marked as (e, that of 35$\pm$4 Bq~kg$^{-1}$ ) and [g, that of O(500 Bq~kg$^{-1}$)] in Fig.~\ref{fig:LabCGePicture}.
The newly purchased 6N grade copper and lead bricks with a ${}^{210}$Pb activity of 5$\pm$3~Bq~kg$^{-1}$ underwent a cleaning process involving immersion in 4\% dilute nitric acid for 20~minutes, followed by rinsing with ultra-pure water for an additional 20~minutes to remove surface contamination. These materials were then wiped with a clean cloth and dried in a chamber supplied with nitrogen gas flowing at a rate of 5 standard liters per minute for a day. Other copper and lead bricks were cleaned with ethanol prior to installation.
As for the cleaning process for 6N copper and lead bricks with a ${}^{210}$Pb activity of 5$\pm$3 Bq~kg$^{-1}$ used in Ge01 shield layers, they were only cleaned with ethanol using an ultrasonic cleaning machine.
Typically, normal samples are placed on a 5~mm thick acrylic stage. The distance between the top of the acrylic stage and the end cap of the Ge detector is 9~mm. 
However, for measuring a large amount of sample, such as the 10~kg Gd$_2$(SO$_4$)$_3{\cdot}$8H$_2$O mentioned in Sect.~\ref{sec:Analysis}, the acrylic stage can be removed, allowing the sample piece to be placed next to the Ge detector.

\subsection{Data acquisition system}
\begin{figure}[htbp]
\begin{center}
\includegraphics[scale = 0.4]{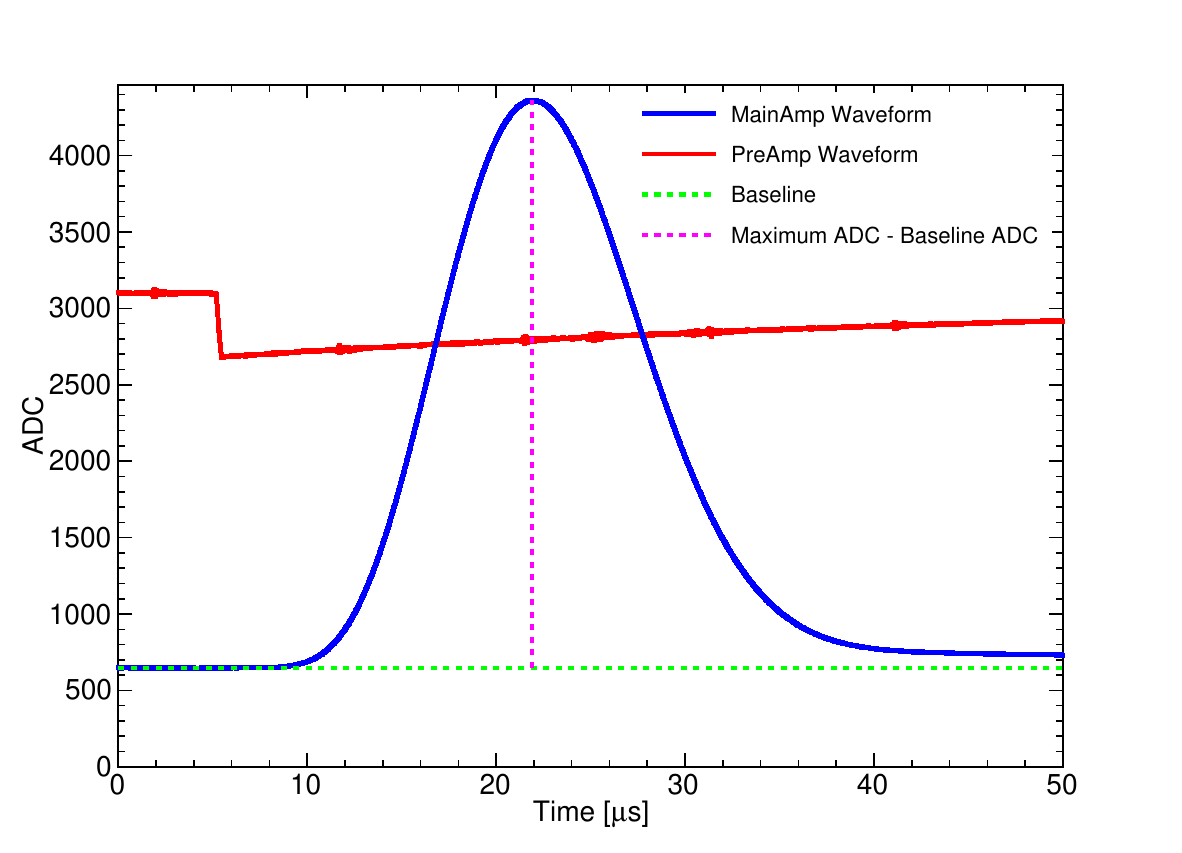}
\caption{Typical preamplifier and main amplifier waveforms for a ${}^{137}$Cs 662~keV photopeak event. The hosizontal dashed line represents the baseline level calculated from the first 2~$\mu$s of the waveform.
The vertical dashed line corresponds to the amplitude of the maximum peak observed in the output signal of the main amplifier. 
This value is obtained by subtracting the baseline ADC from the peak height and is used to estimate the energy associated with the detected gamma-ray event.}
\label{fig:CsWaveform}
\end{center}
\end{figure}

The signal from the Ge02 detector preamplifier was fed into a Canberra Model 2026 main amplifier.
The signal from both the preamplifier and the main amplifier was recorded by a CAEN DT5724B waveform digitizer with a sampling rate of 100~MHz.
The coarse and fine gain settings of the main amplifier were adjusted to avoid saturation of the Analog-to-Digital Converter (ADC) of the waveform digitizer  by the ${}^{208}$Tl 2614~keV photopeak.
Fig.~\ref{fig:CsWaveform} shows the typical waveform of the $^{137}$Cs 662~keV photopeak event. 
The threshold of the main amplifier waveform is 30~ADC counts above the baseline, corresponding to an energy threshold of about 10~keV.
For energy estimation, the maximum amplitude of the pulse from the main amplifier and a baseline calculated from the first 2~$\mu$s of the waveform were used.
Pulse information such as the baseline fluctuation and the channel at the maximum amplitude facilitates the identification and rejection of unphysical electric noise events and pile-up events.

\subsection{Calibration measurement and efficiency determination}
\begin{figure}[htbp]
\includegraphics[width=\linewidth]
{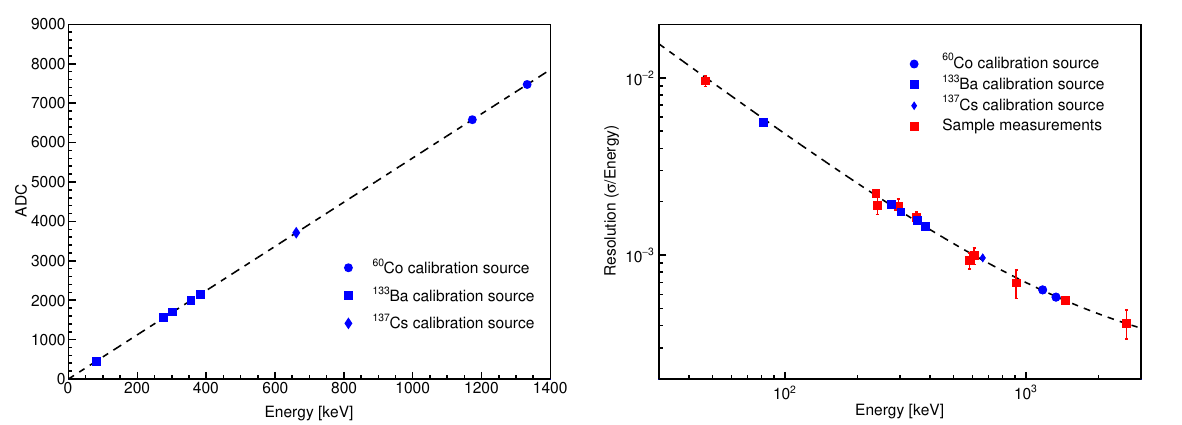}
\caption{(Left) Linearity of the energy response and (right) energy resolution, defined as the $\sigma$ divided by the energy, obtained from calibration sources and sample measurements. The dashed curve represents a fitted result of $\sigma^{2}(E) = E^{2}[(4.95\pm1.07)\times10^{-8}]+E[(2.31\pm0.24)\times10^{-4}]+[(2.09\pm0.10)\times10^{-1}]$ using calibration sources.}
\label{fig:EnergyConversionAndResoluton}
\end{figure}

Detector calibrations have been performed regularly using radioactive sources of ${}^{60}$Co, ${}^{133}$Ba, and ${}^{137}$Cs.
Fig.~\ref{fig:EnergyConversionAndResoluton} shows the linearity of the energy response and the energy resolution as a function of the energy.
Energy is calibrated within 0.06\% uncertainty.
The full width at half maximum (FWHM) of the 1333~keV photopeak of ${}^{60}$Co was measured to be 1.82~keV.
In the comparison between the energy spectra of the experimental data and the Monte Carlo (MC) simulations, the energy resolution function shown in Fig.~\ref{fig:EnergyConversionAndResoluton} (right) is used.

\begin{figure}[htbp]
\includegraphics[width=\linewidth]
{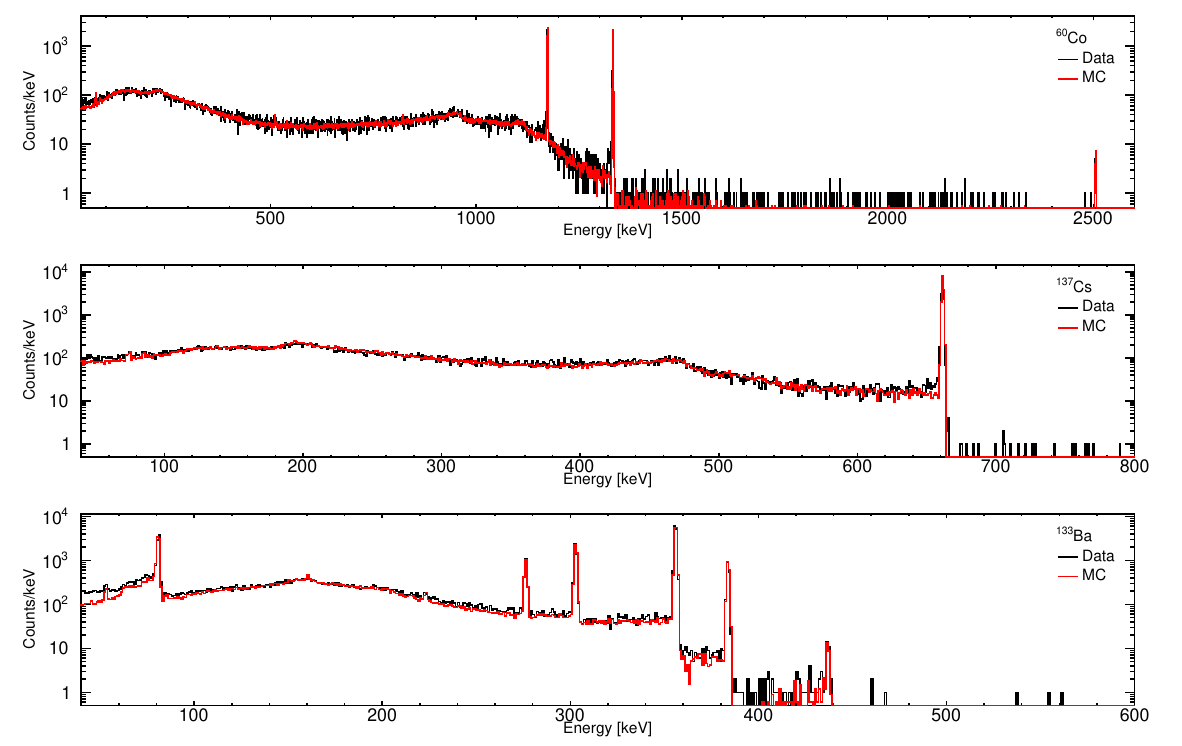}
\caption{Comparison of the energy spectrum from ${}^{60}$Co (top), ${}^{137}$Cs (middle), and ${}^{133}$Ba (bottom) calibration data with that of a Monte Carlo simulation. In the simulation, the energy spectrum is normalized by the measurement time and the activity of the source, and the energy deposition in the HPGe detector is smeared with the energy resolution function described in Fig.~\ref{fig:EnergyConversionAndResoluton}. The differences of the count rate (MC~$-$~Data)/Data in the $\pm$3$\sigma$ region, subtracting the continuous count rate derived from the sideband region, are $-$16.5$\pm$6.9\% (${}^{133}$Ba, 53~keV), 7.6$\pm$1.1\% (${}^{133}$Ba, 81~keV), 1.8$\pm$2.3\% (${}^{133}$Ba, 277~keV), 5.1$\pm$1.4\% (${}^{133}$Ba, 303~keV), 4.5$\pm$0.8\% (${}^{133}$Ba, 356~keV), 8.3$\pm$2.1\% (${}^{133}$Ba, 384~keV), 11.4$\pm$0.9\% (${}^{137}$Cs, 662~keV), 9.5$\pm$1.3\% (${}^{60}$Co, 1173~keV), 9.7$\pm$1.3\% (${}^{60}$Co, 1333~keV), respectively. Here error values correspond to the statistical uncertainties.}
\label{fig:CalibDataMCComparison}
\end{figure}

The MC simulations based on Geant4~\cite{AGOSTINELLI2003250} are performed to compute the detection efficiencies for different gamma-lines used to assess radioactivities, as described in Sect.~\ref{sec:SampleMeasurements}. 
These simulations take into account the detailed geometry of the detector, including the shielding layers.
Fig.~\ref{fig:CalibDataMCComparison} shows a comparison between the spectra obtained from the ${}^{60}$Co, ${}^{133}$Ba, and ${}^{137}$Cs calibrations and those obtained from the simulations which trace the full decay chain of the isotopes using G4RadioactiveDecay.
Except for an unknown continuous component excess of data below 60~keV, the simulation reproduces the calibration data well.
The peak count rates observed in the calibration data, with the source placed at different positions, agree within 10\% with those obtained from the MC simulation. 
Following the approach outlined in Refs.~\cite{ABE2019171,Abe_2020}, since the systematic uncertainties associated with the detection efficiencies derived from the simulation are common to all measurements, the results presented in Sect.~\ref{sec:SampleMeasurements} are given only with their respective statistical uncertainties. 

\section{Data analysis methods}
\label{sec:Analysis}
\subsection{Background spectra}
\begin{figure}[ht]
\includegraphics[width=\linewidth]
{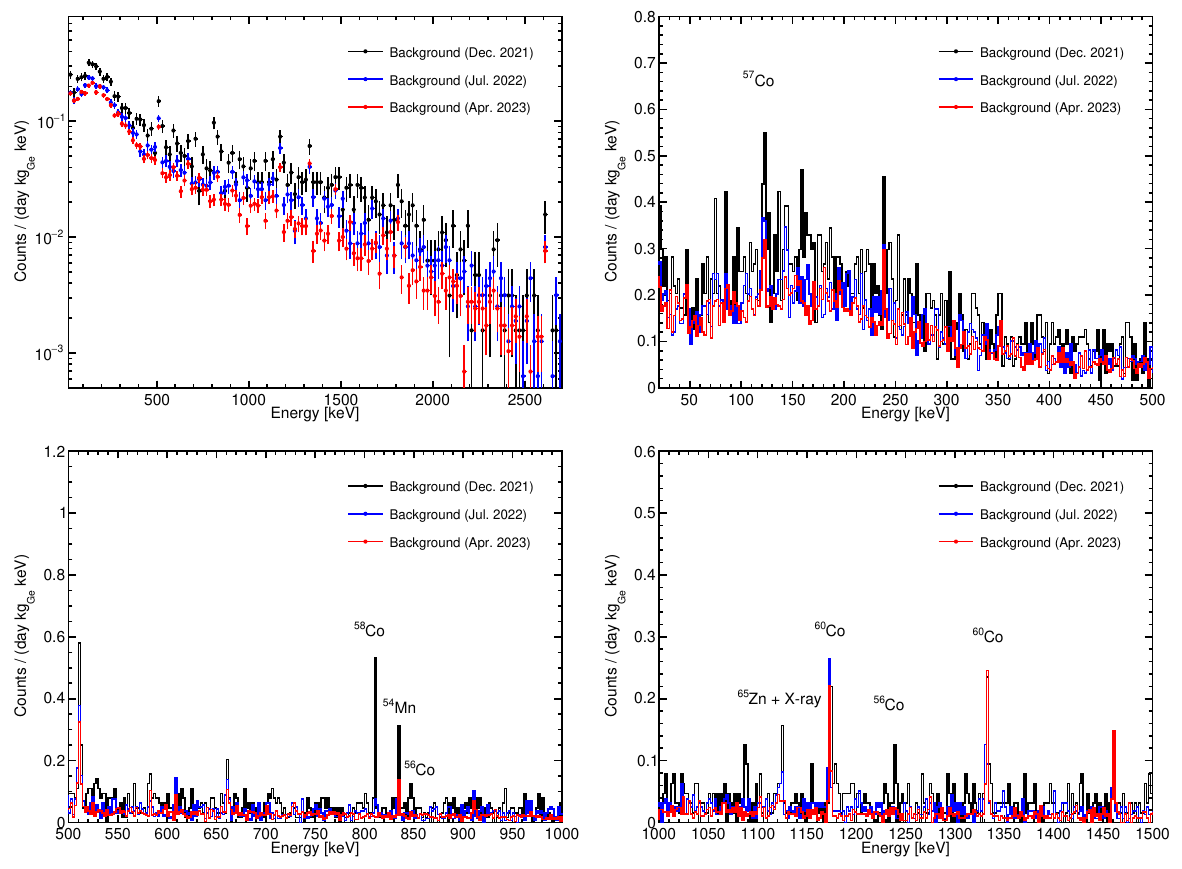}
\caption{The energy spectrum of the background data taken on December 2021, July 2022, and April 2023 for the energy ranges of [20,2700]~keV (top left, 20~keV bin, with statistical error bar), [20, 500]~keV (top right), [500, 1000]~keV (bottom left), and [1000,1500]~keV (bottom right). The bin width for the linear-scale figures is 2~keV. Clear peaks induced by cosmogenic activation in germanium and copper are listed in the linear-scale figures.}
\label{fig:BGDataComparison}
\end{figure}

Data acquisition with the shielding described in Sect.~\ref{sec:DetectorConfiguration} started in December 2021. 
During the sample measurements described in Sect.~\ref{sec:SampleMeasurements}, three long-term BG data acquisitions were performed.
Fig.~\ref{fig:BGDataComparison} shows the time variation of the BG spectra.
The continuous component and the distinct peaks due to cosmogenic activation in germanium and copper, as mentioned in Ref.~\cite{CEBRIAN2010316}, such as ${}^{57}$Co (122~keV, T$_{1/2}$ = 271.74~d) and ${}^{58}$Co (811~keV, T$_{1/2}$ = 70.86~d), show a decreasing trend with time.
The peaks attributed to ${}^{210}$Pb and ${}^{137}$Cs, which were observed in the energy spectra of Ge01 (Fig.~6 in Ref.~\cite{10.1093/ptep/ptac170}), also show a significant decrease from Ge01.

Table~\ref{tab:Gedetectors} shows the BG count rate for the integral in the energy range from 40 to 2700~keV, as well as for several gamma-lines.
After the decrease of radioactivity due to cosmogenic activation, the integral count rate in the energy range from 40 keV to 2700 keV of Ge02 becomes 25\% lower than that of Ge01. 
Table~\ref{tab:GedetectorsComparison} summarizes the crystal mass, energy resolution and BG count rate in comparison with other ultra-low BG HPGe detectors used for material screening at the other experimental site. The BG level of Ge02 is comparable to that of the HPGe detectors used for material screening at the most advanced experiments.

\begin{table}[htbp]
\centering
\caption{The time variation of the background count rates for the integral in the 40--2700~keV region and integral count rates in the $\pm$3$\sigma$ region for several gamma-lines. For the Ge01 detector, the count rates are calculated from the background data used in Ref.~\cite{10.1093/ptep/ptac170}.}

\small
\begin{tabular}{c|c|ccc}
\hline\hline   
Detector   & Ge01 &\multicolumn{3}{c}{Ge02}\\ \hline
Date & Dec. 2019 & Dec. 2021 & Jul. 2022 & Apr. 2023 \\
Measurement time (d) & 23.0 & 19.0 & 47.2 & 86.2 \\ \hline
\multicolumn{5}{c}{Count rate (kg$_{Ge}^{-1}$ day$^{-1}$)}\\ \hline
Integral 40 -- 2700~keV & 112.6 & 140.2 & 100.0 & 84.3 \\  
$^{208}$Tl, 2614~keV & 0.08$\pm$0.04 & 0.25$\pm$0.09 & 0.16$\pm$0.05 & 0.13$\pm$0.03 \\
$^{214}$Bi, 609~keV & 0.39$\pm$0.10 &  0.25$\pm$0.09 & 0.38$\pm$0.07 & 0.23$\pm$0.04 \\
$^{60}$Co, 1333~keV & 0.41$\pm$0.10 &  0.66$\pm$0.14 & 0.48$\pm$0.08 & 0.68$\pm$0.07 \\
$^{40}$K, 1461~keV & 0.44$\pm$0.11 &  0.31$\pm$0.10 & 0.44$\pm$0.07 & 0.42$\pm$0.05 \\
$^{137}$Cs, 662~keV & 1.29$\pm$0.18 &  0.53$\pm$0.13 & 0.38$\pm$0.07 & 0.32$\pm$0.05 \\
$^{210}$Pb, 46.5~keV & 3.24$\pm$0.29 &  0.69$\pm$0.14 & 0.64$\pm$0.09 & 0.59$\pm$0.06 \\ \hline
\end{tabular}
\label{tab:Gedetectors}
\end{table}

\begin{table}[htbp]
\centering
\caption{Comparison of crystal mass, relative efficiency, energy resolution and background rate with other HPGe detectors used for material screening at the other underground experimental site. The HPGe detectors described in Refs.~\cite{10.1093/ptep/ptac170, Araujo_2022} are listed.
}

\small
\begin{tabular}{c|ccccc}
\hline\hline   
 \multirow{3}{*}{Site} & \multirow{3}{*}{Detector} & Crystal  & Relative & FWHM at & BG rate \\
 & & mass &  efficiency & 1333 keV & 60 -- 2700 keV \\
 & &  [kg] & [\%] &  [keV] & [kg$_{Ge}^{-1}$ day$^{-1}$)] \\ \hline
 \multirow{2}{*}{Kamioka} & Ge02 (This work) & 1.68 & 80 & 1.82 & 81.3$\pm$0.7 \\
 & Ge01~\cite{10.1093/ptep/ptac170} & 1.68 & 80 & 2.39 &  104.5 \\ \hline
 \multirow{2}{*}{LNGS} & Gator~\cite{Araujo_2022} & 2.2 & 100.5 & 1.98 & 89.0$\pm$0.7 \\
   & GeMPI~\cite{Araujo_2022} & 2.2 & 98.7 & 2.20 & 24$\pm$1 \\ \hline
\multirow{2}{*}{BUGS} & Belmont~\cite{10.1093/ptep/ptac170} & 3.2 & 160 & 1.92 & 90.0 \\
   & Merrybent~\cite{10.1093/ptep/ptac170} & 2.0 & 100 & 1.87 & 145.0 \\ \hline
\multirow{3}{*}{LSC} & GeOroel~\cite{10.1093/ptep/ptac170} & 2.31 & 109 & 2.22 & 128.7 \\
& Asterix~\cite{10.1093/ptep/ptac170} & 2.13 & 95.1 & 1.92 & 171.3 \\
& GeAnayet~\cite{10.1093/ptep/ptac170} & 2.26 & 109 & 1.99 & 461.2 \\ \hline
BHUC & Maeve~\cite{LZ_Screening} & 2.0 & 85 & 3.19 & 956.1 \\ \hline
LVdA & GeMSE~\cite{Araujo_2022, GeMSE} & 2.0 & 107.7 & 1.96 & 88$\pm$1 \\ \hline
\end{tabular}
\label{tab:GedetectorsComparison}
\end{table}

\subsection{Sample measurements}
\label{sec:SampleMeasurements}
\begin{figure}[htbp]
\includegraphics[width = \linewidth]
{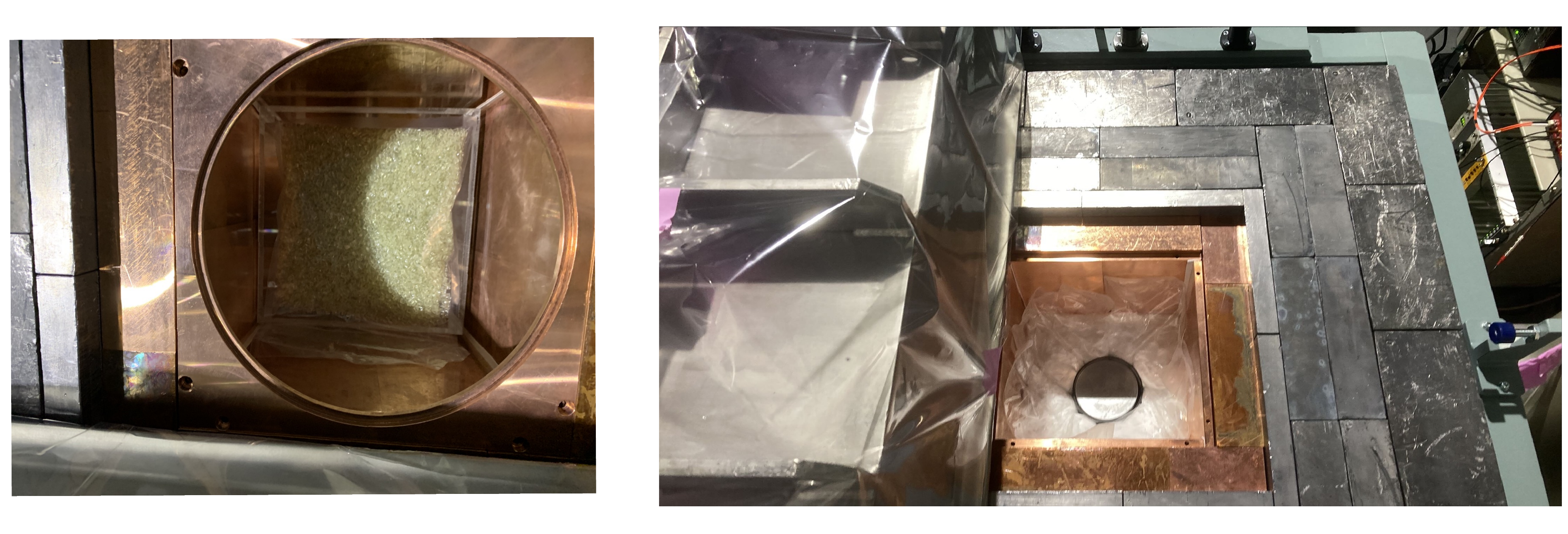}
\caption{Pictures of a 440~g PEN pellet sample on the acrylic sample stage(left) and Ge02 detector with 10~kg of Gd$_{2}$(SO$_{4}$)$_{3}$$\cdot$8H$_{2}$O sample (right).}
\label{fig:SamplePictures}
\end{figure}

Samples are flushed with Rn-free air, evacuated, and sealed in ethylene-vinyl alcohol (EVOH) bags.
EVOH bags retain Rn emanated from samples.
Fig.~\ref{fig:SamplePictures} shows  photographs of a 440~g PEN pellet sample and a 10~kg of Gd$_2$(SO$_4$)$_3{\cdot}$8H$_2$O sample.

\begin{figure}[htbp]
\includegraphics[width = \linewidth]
{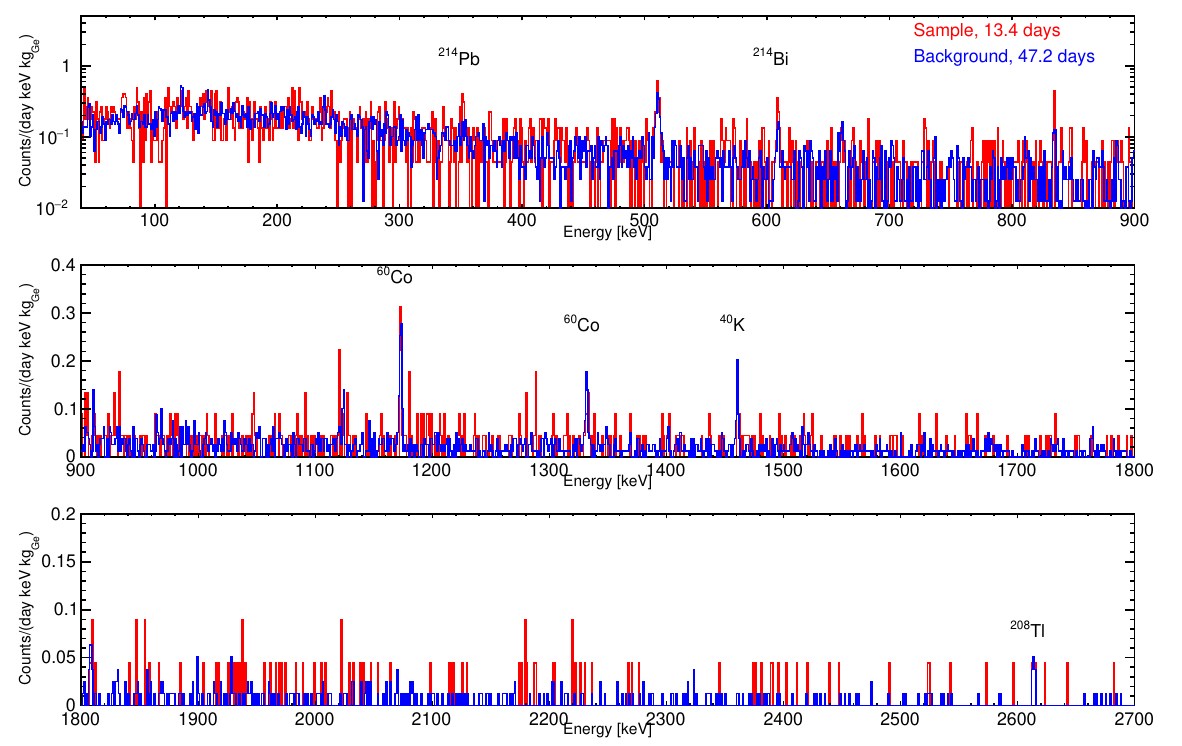}
\caption{The energy spectra for the PEN pellet sample measurement and the background.}
\label{fig:PENEnergySpectrum}
\end{figure}

\begin{figure}[htbp]
\includegraphics[width = \linewidth]
{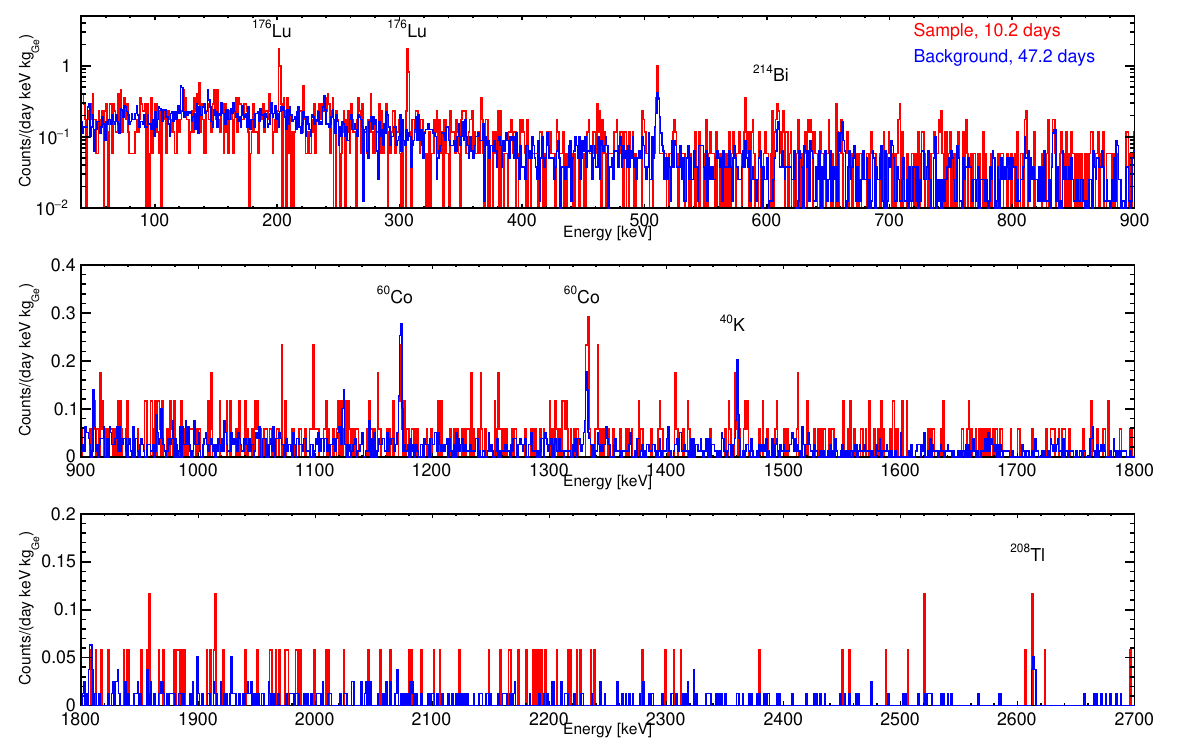}
\caption{The energy spectra for a typical 10 kg of Gd$_2$(SO$_4$)$_3{\cdot}$8H$_2$O sample measurement and the background.}
\label{fig:Gd2SO4_3EnergySpectrum}
\end{figure}

 Data collected within the first 12~hours from sample deployment were not used in  the analysis to mitigate the effects of Rn which was contaminated into the chamber during the sample deployment.
 A noisy period associated with the supply of liquid nitrogen was also excluded from the analysis.
Figs.~\ref{fig:PENEnergySpectrum} and~\ref{fig:Gd2SO4_3EnergySpectrum} show the energy spectra of the sample and the BG measurements.

The determination of the radioactivity of the sample is based on the net signal count $S_{net}$, which is defined as:
\begin{equation}
S_{net}  = S_{peak} - S_{comp} - (B_{peak} - B_{comp})\cdot T_{s}/T_{b},
\end{equation}
Here, $S_{peak}$ represents the counts in the photopeak region. The photopeak region is defined as the $\pm$4 keV region from the peak, that is sufficiently larger than the energy scale uncertainty and $\pm$3$\sigma$ of the energy resolution. 
$S_{comp}$ corresponds to the Compton BG counts and is estimated from the $\pm$4 keV regions adjacent to the photopeak region. $B_{peak}$ and $B_{comp}$ denote the counts in the photopeak and Compton regions of the BG data, respectively. $T_s$ and $T_b$ are the measurement times for the sample and the BG, respectively.

The specific activity A~[Bq~kg$^{-1}$] for a given gamma-line is determined using the equation:
\begin{equation}
A ~{\rm[Bq~kg^{-1}]} = \frac{S_{net}}{I_{\gamma}\cdot\epsilon\cdot m\cdot T_{s}}.
\end{equation}
In this equation, $I_{\gamma}$ represents the branching ratio of the gamma-ray, $\epsilon$ is the photopeak detection efficiency and $m$ is the mass of the sample in kilograms. The term $\epsilon$ is evaluated by the MC simulation using regular-shaped geometry closest to that of the sample. The density and composition of the sample are also taken into account by the MC to consider self-absorption in the samples. The simulated monochromatic gamma-rays are generated uniformly and isotropically within the sample volume. 
If several prominent gamma-lines originate from a specific radioisotope~(RI), the radioactivity is determined using the weighted average of the evaluations for each gamma-line. For example, gamma-lines from ${}^{214}$Pb (295~keV) and ${}^{214}$Bi (609, 1120, and 1764~keV) are used to assess the radioactivity of ${}^{226}$Ra.
The effect of the coincidence sum of multiple gamma-rays in the decay cascade was simulated for these samples and found to be negligible compared to the statistical error.
Following Refs.~\cite{ABE2019171,Abe_2020}, if the central value does not exceed twice the statistical error or is negative, the result is considered to be consistent with zero. In such cases, an upper limit at the 90\% confidence level (C.L.) is calculated as max\{0, the central value\} + 1.28 $\times$ the statistical error.
Table~\ref{tab:RIResults} shows the measured RIs for the PEN pellet and Gd$_2$(SO$_4$)$_3{\cdot}$8H$_2$O samples.
With the exception of ${}^{176}$Lu in the Gd$_2$(SO$_4$)$_3{\cdot}$8H$_2$O sample, no significant RIs were observed. 
Similar to Ge01, Ge02 demonstrates sensitivity to less than 0.5~mBq~kg$^{-1}$ of ${}^{226}$Ra in the Gd$_2$(SO$_4$)$_3{\cdot}$8H$_2$O sample at a 90\% C.L. with about 10~days of measurement.
\begin{table}[htbp]
\small
    \centering
    \caption{Summary of the measurement results [mBq~kg$^{-1}$] for PEN pellet and Gd$_2$(SO$_4$)$_3{\cdot}$8H$_2$O sample. Upper limits correspond to the 90\% confidence level. }
    \begin{tabular}{lcc}
    \toprule
    Sample & PEN Pellet  & Gd$_2$(SO$_4$)$_3{\cdot}$8H$_2$O \\
    Measurement time [d] & 13.4 & 10.2 \\
    \midrule
    Early part of ${}^{238}$U chain & \multirow{3}{*}{$<$ 17.4} & \multirow{3}{*}{$<$ 6.2} \\
     ${}^{234}$Th 63, 93~keV &  & \\
     ${}^{234\mathrm{m}}$Pa 1001 keV & & \\
     \midrule
    middle part of ${}^{238}$U chain & \multirow{3}{*}{$<$ 2.60} & \multirow{3}{*}{$<$ 0.43} \\
     ${}^{214}$Pb 295~keV & & \\
     ${}^{214}$Bi 609, 1120, 1764~keV & & \\
    \midrule
    late part of ${}^{238}$U chain & \multirow{2}{*}{$<$ 137} & \multirow{2}{*}{$<$ 91} \\
     ${}^{210}$Pb 46.5~keV & & \\
    \midrule
    Early part of ${}^{232}$Th chain & \multirow{2}{*}{$<$ 2.10} & \multirow{2}{*}{$<$ 0.25} \\
     ${}^{228}$Ac 338, 911, 969~keV &  & \\
    \midrule
    late part of ${}^{232}$Th chain & \multirow{3}{*}{$<$ 1.96} & \multirow{3}{*}{$<$ 0.16} \\
     ${}^{212}$Bi 727~keV &  & \\
     ${}^{208}$Tl 583, 861, 2614~keV &  & \\
    \midrule
    Early part of ${}^{235}$U chain & \multirow{2}{*}{$<$ 5.56} & \multirow{2}{*}{$<$ 1.79} \\
     ${}^{235}$U 144, 163, 205~keV &  & \\
    \midrule
    late part of ${}^{235}$U chain & \multirow{4}{*}{$<$ 3.47} & \multirow{4}{*}{$<$ 0.66} \\
    ${}^{223}$Ra 154, 269~keV &  & \\
    ${}^{219}$Rn 271, 402~keV &  & \\
    ${}^{211}$Pb 405, 427, 832~keV &  & \\
    \midrule
    ${}^{40}$K 1461~keV & $<$ 7.63 & $<$ 0.86 \\
    \midrule
    ${}^{60}$Co 1173, 1333~keV & $<$ 0.66 & $<$ 0.07 \\
    \midrule
    ${}^{176}$Lu 202, 307~keV & $<$ 0.52 & 0.46 $\pm$ 0.09 \\
    \bottomrule
    \end{tabular}
\label{tab:RIResults}
\end{table}

Almost all Gd$_2$(SO$_4$)$_3{\cdot}$8H$_2$O lots used in the second Gd-loading to Super-Kamiokande were screened with the Ge01 and Ge02 detectors. 
Further details on the screening of Gd$_2$(SO$_4$)$_3{\cdot}$8H$_2$O will be found in K. Abe~{\it et al.}~[manuscript in preparation].

\section{Conclusion}
\label{sec:Conclusion}
We have successfully deployed a new low-background HPGe detector at the Kamioka Observatory. 
The integral background count rate in the energy range from 40 keV to 2700 keV is about 25\% lower than that of Ge01 and comparable to that of the HPGe detectors used for material screening at the most advanced experiments.
The HPGe detector demonstrates a sensitivity of less than 0.5~mBq~kg$^{-1}$ of ${}^{226}$Ra in a 10~kg Gd$_2$(SO$_4$)$_3{\cdot}$8H$_2$O sample at a 90\%  C.L. with an approximate measurement time of 10~days. 
This detector is currently playing essential roles for material screenings for the SK-Gd, KamLAND-Zen, and other research and development experiments.

\section*{Acknowledgment}
We thank Mirion Technologies staff for fruitful discussions and the development of an ultra-low BG HPGe detector. We would like to thank Toyobo Co., Ltd. and Nippon Yttrium Co., Ltd. for  providing the samples. We thank the technical staffs of the RCNS and ICRR for their assistance with the detector installation work.
This work was supported by the Japanese Ministry of Education, Culture, Sports, Science and Technology, Grant-in-Aid for Scientific Research, JSPS KAKENHI Grant No. 19H05807, 19H05808, 21H01105, the joint research program of the Institute for Cosmic Ray Research (ICRR), the University of Tokyo.

\bibliographystyle{ptephy2}
\bibliography{Reference}

\end{document}